\documentclass[sigplan,screen]{acmart} 

\AtBeginDocument{%
  }

\usepackage{subcaption}
\usepackage{microtype}

\setcopyright{rightsretained}
\acmPrice{}
\acmDOI{10.1145/3623476.3623528}
\acmYear{2023}
\copyrightyear{2023}
\acmSubmissionID{sle23main-p91-p}
\acmISBN{979-8-4007-0396-6/23/10}
\acmConference[SLE '23]{Proceedings of the 16th ACM SIGPLAN International Conference on Software Language Engineering}{October 23--24, 2023}{Cascais, Portugal}
\acmBooktitle{Proceedings of the 16th ACM SIGPLAN International Conference on Software Language Engineering (SLE '23), October 23--24, 2023, Cascais, Portugal}
\received{2023-07-07}
\received[accepted]{2023-09-01}

\begin{document}


\title[Online Name-Based Navigation for Software Meta-languages]{Online Name-Based Navigation for \\ Software Meta-languages}


\author{Peter D. Mosses}
\orcid{0000-0002-5826-7520}
\email{P.D.Mosses@tudelft.nl}
\affiliation{%
  \institution{TU Delft}
  \city{}
  \country{Netherlands}
  \\
  \institution{Swansea University}
  \city{}
  \country{UK}
}

\begin{abstract}

Software language design and implementation often involve specifications written in various esoteric meta-languages. 
Language workbenches generally include support for precise name-based navigation when browsing language specifications \emph{locally}, but such support is lacking when browsing the same specifications \emph{online} in code repositories.

This paper presents a technique to support precise name-based navigation of language specifications in online repositories using ordinary web browsers.
The idea is to generate \emph{hyperlinked twins}: websites where \emph{verbatim copies} of specification text are enhanced with hyperlinks between name references and declarations.
By generating hyperlinks directly from the name binding analysis used internally in a language workbench, online navigation in hyperlinked twins is automatically consistent with local navigation.

The presented technique has been implemented for the Spoofax language workbench,
and used to generate hyperlinked twin websites from various language specifications in Spoofax meta-languages.
However, the applicability of the technique is not limited to Spoofax, and developers of other language workbenches could presumably implement similar tooling, to make their language specifications more accessible to those who do not have the workbench installed.

\end{abstract}




\maketitle

\section{Introduction}
\label{sec:introduction}

Name-based navigation is a significant aspect of software language engineering.
IDEs generally include support for precise name-based navigation when browsing code \emph{locally},
but such support is lacking \emph{online} when using ordinary web-browsers on code repositories.

Here, we suggest to generate \emph{hyperlinked twin websites} from code repositories.
The code on the website should look the same as it does in an IDE,
and the hyperlinks should support the same name-based navigation as the IDE.

Software \emph{meta-languages} are a particularly important special case of software languages,
and language workbenches implement name-based navigation for the meta-languages that they use.
Moreover, a language workbench is likely to provide an API to access ASTs and name binding analyses,
facilitating generation of hyperlinked twin websites.

To illustrate the suggested technique, the Spoofax language workbench \cite{Kats2010SLW} has been used
to generate hyperlinked twins from various language specifications in Spoofax meta-languages.%
\footnote{\url{https://pdmosses.github.io/hyperlinked-twins/}}
This involved writing only a small amount of code in the Spoofax meta-language Stratego.
The code uses generic AST traversals to generate HTML from parsed and analysed specifications,
and a simple API for accessing name binding information.
The code is available on GitHub.%
\footnote{\url{https://github.com/pdmosses/sdf/tree/master/org.metaborg.meta.lang.template/trans/generation/docs/}}

The rest of this section expands on the above points.
Section~\ref{sec:generation} then explains the main steps of the generation process,
which may be of interest to developers of other language workbenches.
Section~\ref{sec:spoofax} briefly mentions some details specific to the use of Spoofax.
Section~\ref{sec:conclusion} concludes, and discusses future work.
Appendix~\ref{sec:screenshots} shows how a fragment of a language specification looks
in Spoofax, in a GitHub repository, and in the hyperlinked twin generated from that repository.

\subsection{Name-Based Code Navigation}

Software languages generally include \emph{declarations} that bind names to entities,
and \emph{references} to those entities using the declared names.
Name-based navigation between declarations and references is essential for browsing and exploring code in software languages.

Manual name-based navigation can be tedious and error-prone: it may require scrolling, or entering text in search boxes.
It becomes significantly more difficult when declarations can be in different files from references to them –
particularly when code is divided into hundreds of files,
perhaps with a complicated import relationship.

Integrated software development environments (IDEs) support name-based navigation when locally browsing or editing code.
When a reference to a name is selected, the IDE allows navigation directly to the relevant declaration(s).
When a declaration is selected, the IDE may also support navigation directly to some or all the references to it.

Often, a name can be used in more than one declaration in the same project –
either in different namespaces (e.g., types and constructors) or in different parts of the project.
Support for name-based navigation using simple textual search may then be significantly inferior
to precise navigation using name binding analysis,
due to false positives in search results.

Support for name-based navigation is often weak in online code repositories when using ordinary web browsers.
GitHub repositories currently support search-based code navigation in about a dozen mainstream programming languages \cite{GitHub-ANC},
but precise name-based navigation in only one language  \cite{GitHub-PCN}: Python.
GitHub's implementation of precise online name-based navigation 
requires specifying the name binding analysis of the language
in terms of stack graphs \cite{Creager2023SG}.
Apart from the significant amount of expertise and effort required for that,
a potential drawback of GitHub's approach may be the difficulty of validating that the navigation in the repository
accurately reflects the name-binding analysis implemented in compilers.
In any case, precise navigation on GitHub seems likely to be limited to a few major programming languages,
despite the possibility for language developers to contribute support for further languages \cite{GitHub-BCN}.

\subsection{Software Meta-languages}

A \emph{meta-language} is a language for specifying languages (primarily their syntax and semantics).
A \emph{software meta-language} is simply a meta-language for specifying software languages.
Specifications of major software languages can be large, and difficult to navigate.
Moreover, unfamiliarity with a particular software meta-language can hinder manual name-based navigation
in language specifications –
especially when name binding in the meta-language differs significantly from that in conventional programming languages.

Development and validation of software language specifications is supported by software language workbenches,
which generally implement precise name-based navigation.
However, that navigation is not generally available for such language specifications
when browsing them in online repositories using ordinary web browsers.
To browse a language specification with precise name-based navigation,
users then need to install a workbench locally and download a copy of the repository.

\subsection{Prior Examples of Hyperlinked Twins}

The reference manuals of most current programming languages are available online in HTML or PDF, and can be browsed using ordinary browsers.
There, hyperlinks already support name-based navigation in grammars that specify language syntax.
When the hyperlinks are generated from repositories containing the plain text of the grammars,
the reference manuals may then be regarded as hyperlinked twins.

The author has previously developed support for precise name-based navigation of language specifications online:
the CBS-beta website,%
\footnote{\url{https://plancomps.github.io/CBS-beta/}}
which was generated from CBS specifications whose syntax and name binding were specified in Spoofax meta-languages.
In \cite{Mosses2023USS} he speculated that the approach used to generate the CBS-beta website might be applicable to other software meta-languages;
the present paper confirms that,
but it turned out not to be possible to reuse the implementation of the generation process directly:
the code involved case analysis on the constructs of CBS,
and would need to be almost completely reimplemented for each meta-language. 

Various other specification frameworks provide tool support for generating hyperlinked websites from specifications.
For example, the web version of an online book \cite{Wadler2022PLF} includes hyperlinked pages
generated from (literate) Agda source code.
If web versions of source code in other specification languages can be generated using the same tool support,
it would be interesting to compare the generation process with that outlined here.

\section{Generating Hyperlinked Twin Websites}
\label{sec:generation}

The aim is tool support for online name-based navigation of language specifications in ordinary web browsers. 
The main idea is to generate web pages where verbatim copies of the specifications are enhanced with hyperlinks
between name references and declarations.
By generating the hyperlinks directly from analyses used internally in language workbenches,
online navigation in language specifications is automatically consistent with local navigation.

The proposed technique has been implemented in the Spoofax language workbench, with only modest effort,
as outlined in Section~\ref{sec:spoofax};
it might be possible to implement it in other language workbenches in much the same way.
%

Suppose that some language workbench is to generate a hyperlinked website from the plain code of a language specification found online in some GitHub repository.
The suggested technique is to proceed as follows.

\paragraph{Requirements.}

The language workbench needs to parse and analyse the plain language specification.
Unless the workbench can directly access the repository online, a local clone is required;
and to add the source files for the generated website to the repository using pull-requests,
the clone will need to be published as a fork of the repository.

If the language specification is in meta-languages supported by the workbench,
it can already parse and analyse them.
However, the results also need to be accessible for transformation to HTML.
(That should always be possible when the implementation of the meta-languages in the workbench is bootstrapped.)
If the specification uses external meta-languages, those languages need to be loaded into the workbench before proceeding.

The following steps are to be applied to a complete language specification project.

\paragraph{Creating ASTs.}

To support generation steps that involve tree traversal,
the first step is to parse the language specification files and create corresponding abstract syntax trees (ASTs).
The generation process is to be completely independent of the detailed structure of the ASTs
(and hence of the meta-language used for specification).
The ASTs might correspond closely to parse trees,
or they could be `de-sugared' to remove semantically-irrelevant structure
such as white space, line breaks, and literal terminal symbols
(depending on the language).

However, the ASTs must support the addition of name binding information
to nodes that correspond to declarations and references.
Such nodes also need to reveal the start and end positions of their source text.

The language workbench may automatically parse files and generate their ASTs,
otherwise this step needs to be explicitly executed.

\paragraph{Adding name binding analysis.}

Based on the relevant name binding analysis for the meta-language,
this step should ensure that all declarations and references can be detected when traversing the ASTs.
Each declaration node needs to provide the source text of the declared name;
each reference node needs to provide not only the name,
but also the declaration(s) to which the reference has been resolved.

In general, a reference may resolve to a declaration in a different file;
and a declaration of a single name may be spread across multiple files.

As with generating ASTs, a language workbench may automatically analyse files
and add the resulting information to their ASTs,
otherwise this step needs to be explicitly executed.
The remaining steps are specific to the generation of hyperlinked websites,
but could also be made automatic.

\paragraph{Generating plain HTML.}

The obvious way to generate HTML that renders exactly as some plain source text is to enclose the text
in \verb|<pre><code>...</code></pre>| tags.
In general, this preserves the white space (i.e., indentation and line breaks) of the source text
– assuming that the rendering uses a fixed-width font.

The source text might also contain the characters `\verb|<|', `\verb|>|', and `\verb|&|',
which are all treated specially in HTML.
These need to be replaced by the corresponding HTML entities `\verb|&lt;|', `\verb|&gt;|', and `\verb|&amp;|', respectively.

Subsequent steps are to enclose parts of the source text in tags for hyperlinks and highlighting.
To avoid the need for obtaining the source text of all nodes in an analysed AST,
plain HTML can be generated gradually, 
by copying characters from the source file to the generated file while traversing the AST
(top down, left to right).

\paragraph{Generating hyperlinks.}

To generate hyperlinks between declarations and references,
the relevant tags can be inserted whenever the traversal reaches the corresponding node.

When the node is a declaration of name $N$ at position $P$, the element
\verb|<span id="|$N$\_$P$\verb|">|$N$\verb|</span>|
provides a unique target for references that resolve to this declaration of $N$.
The inclusion of the position $P$ ensures that the ID of the tag is unique in the generated file

Similarly, when a reference to name $N$ resolves to a single declaration of $N$ at position $P$ in file $F$,
the anchor element 
\verb|<a href="|$F$\verb|#|$N$\_$P$\verb|">|$N$\verb|</a>|
renders as the desired hyperlink to the declaration.

In general, a reference to a single name may resolve (unambiguously) to multiple declarations, possibly located in multiple files.
Similarly, multiple references may resolve to the same declaration(s).
Such information can be added to HTML elements as a \verb|title| attribute,
which is usually displayed by HTML browsers as a tooltip while hovering over the element.
(Pop-ups or modals could support links to multiple targets, but might be too distracting
due to the high density of names in language specifications.)

\paragraph{Generating highlighting.}

Independently of name-based navigation, language workbenches use syntax highlighting to enhance code readability.
To make code rendered on the generated website look the same as in a workbench,
the website needs to replicate the colours and fonts that it uses.

Websites often highlight code in many software languages automatically.
For example, GitHub highlights code in its repositories for hundreds of languages,
using Tree-sitter%
\footnote{\url{https://tree-sitter.github.io/tree-sitter/}}
parsing
and context-aware token scanning to recognise different kinds of language construct –
also coping gracefully with incomplete or syntactically ill-formed code.

When a code editor of a language workbench supports the same automatic highlighting framework as a website,
it might seem attractive to exploit it, and avoid the need for adding highlighting markup when generating web pages.
However, this seems incompatible with the simple approach adopted here for generating hyperlinks in HTML\@.
In any case, websites seldom support automatic highlighting for software \emph{meta}-languages.

So here, highlighting is added to generated HTML using tags of the form
\verb|<span class="|$C$\verb|">...</span>|,
where $C$ indicates the (syntactic or lexical) sort of the enclosed text.
The rendering of the text – font colour, style, and weight – can then be specified in CSS (generated from data in the language workbench).

\paragraph{Generating a website.}

When generating a website from code in a repository, it is natural to generate a separate web page for each code file,
and copy the directory structure.
The website navigation panel can then display the directory structure as a tree,
with links to the individual pages as leaves.
The detailed rendering of the navigation panel on the website is not so important,
because name-based navigation reduces (or even eliminates) the need for drilling down through the directory structure of a code project
when browsing or exploring code online.

Static site generators (SSGs) such as MkDocs%
\footnote{\url{https://www.mkdocs.org}}
and Jekyll%
\footnote{\url{https://jekyllrb.com}}
can generate websites automatically from HTML files.
Meta-data can be prefixed to the HTML content as so-called front matter,
e.g., specified in YAML\@.
HTML can also be embedded directly in Markdown,
which facilitates the inclusion of headings and links in the generated source files for the website.
An important advantage of relying on an SSG to generate web pages from Markdown is that the resulting HTML
can be expected to render properly in any (modern) web browser, on mobile devices as well as desktop and laptop computers.

\begin{figure*}
  \footnotesize
\begin{verbatim}
                    <a href="../AssignmentOperators.sdf3#FieldAccess_938_949" id="FieldAccess_331_342"
                  title="Referenced at ../AssignmentOperators.sdf3 line 30; ../Disambiguation.sdf3 line 57;
                  line 16">FieldAccess</a>.<span class="cons_Constructor"><span id="QSuperField_343_354"
                  title="Not referenced locally, nor via imports">QSuperField</span></span> = &lt;&lt;<a
                  href="../../names/Names.sdf3#TypeName_145_153" id="TypeName_359_367" title="Defined
                  at ../../names/Names.sdf3 line 11, 21, 22">TypeName</a>&gt;<span class="cons_String">
                  .super.</span>&lt;<a href="../../lexical/Identifiers.sdf3#Id_141_143" id="Id_376_378"
                  title="Defined at ../../lexical/Identifiers.sdf3 line 15, 23">Id</a>&gt;&gt;
  \end{verbatim}

  \Description{Illustrates the form of generated source files}
  \caption{A fragment of a generated source file for a hyperlinked twin.}
  \label{fig:html}
\end{figure*}

Figure~\ref{fig:html} illustrates the form of the generated HTML. It is a single line from a source file for a hyperlinked twin website (here wrapped to fit the page width).

\section{Using Spoofax}
\label{sec:spoofax}

The Spoofax Language Workbench%
\footnote{\url{https://spoofax.dev}}
currently uses three main meta-languages: SDF3 for syntax, Statix for name binding, and Stratego for transformation.
The meta-languages are themselves specified using Spoofax meta-languages
(including the now-deprecated SDF2, NaBL, and NaBL2).
A further meta-language is ESV, for specifying editor services, including syntax highlighting details.
The specifications of all the meta-languages are available as Spoofax language projects
on GitHub in repositories of the MetaBorg organisation.%
\footnote{\url{https://spoofax.dev/references/}}

The Spoofax language workbench is implemented as an Eclipse plugin.
To implement generation of hyperlinked websites for an external language specified using Spoofax meta-languages,
it is possible to add the required code to the language specification using the plugin.
(That is how the CBS-beta website was generated,
based on the specifications of the CBS meta-language in SDF3 and NaBL2.)

To add the required code to a Spoofax meta-language such as SDF3, however,
it is necessary to build the complete baseline version for bootstrapping Spoofax-2,
following the steps explained in the documentation on Spoofax Development.%
\footnote{\url{https://spoofax.dev/howtos/development/}}
By adjusting the version number in the dependency specification of the relevant meta-language,
Spoofax can be used to parse, analyse, and transform its own specifications.

Spoofax provides a Stratego API for reading text from a file, and for parsing it to produce an AST.
The parser is generated automatically from the SDF3 specification of the language when the language project is built.
The API also supports analysing the name binding of all the files in an Eclipse project,
and adding the analysis as annotations on the AST nodes, which can also be accessed using Stratego.
And it supports accessing the source text of nodes in the AST, which is based on origin-tracking.
The same API includes strategies for obtaining the character positions of name declarations and references.

The generation of a web page with hyperlinks from each source file in a project is specified as a generic traversal in Stratego, independently of the syntax of the language.

\begin{figure*}
  \includegraphics[width=0.8\textwidth]{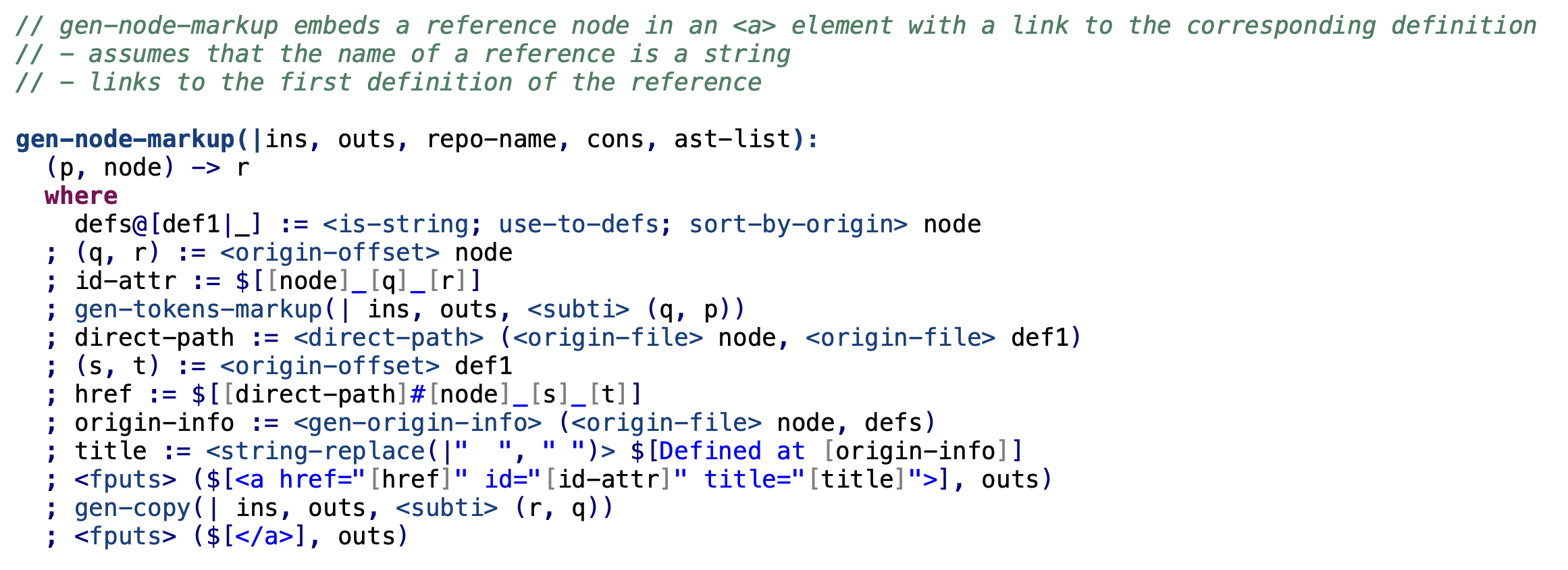}
  \Description{Illustrates Stratego code used for generating HTML}
  \caption{Stratego code for generating HTML from references.}
  \label{fig:stratego}
\end{figure*}

For example, Figure~\ref{fig:stratego} shows the Stratego code for generating HTML from references.

Currently, there is no Stratego API for accessing the kinds of individual lexical tokens determined by parsing.
As a workaround, highlighting markup is added using pattern matches on the source text (expressed by Stratego strategy combinators) and rendered using CSS generated from an ESV specification.
The result corresponds closely to the highlighting in Spoofax.

The documentation site theme used for the main Spoofax documentation website (Material for MkDocs%
\footnote{\url{https://squidfunk.github.io/mkdocs-material/}})
automatically generates a navigation panel with the same structure as the source project,
with language-independent configuration.
However, the underlying MkDocs SSG transforms directory names;
a plugin%
\footnote{\url{https://github.com/lukasgeiter/mkdocs-awesome-pages-plugin}}
is required to ensure that the rendered links in the navigation panel show the untransformed names.

It is straightforward to deploy the generated web pages to GitHub Pages using Actions.
Versioned web pages could also be deployed for different releases or branches.%
\footnote{\url{https://squidfunk.github.io/mkdocs-material/setup/setting-up-versioning/}}

\section{Conclusion and Future Work}
\label{sec:conclusion}

Using the technique presented in this paper, hyperlinked twin websites have been successfully generated from the syntax of several Spoofax meta-languages (SDF3, NABL, NaBL2, Statix) and from the name binding specification of NaBL.%
\footnote{\url{https://pdmosses.github.io/hyperlinked-twins/}}
A future release of Spoofax should support generation of hyperlinked twins from code in all the Spoofax meta-languages, so hyperlinked twins can be published for all repositories that use Spoofax language specifications.
It may also be possible to support meta-languages used in other frameworks.

\begin{acks}
Gabri\"el Konat, Dani\"el Pelsmaeker, and Jeff Smits provided crucial assistance for developing generation of hyperlinked twins using Spoofax.
The comments and suggestions of the anonymous reviewers helped improve the paper.
\end{acks}


\bibliographystyle{ACM-Reference-Format}
\bibliography{SLE}

\appendix

\section{Appendix}
\label{sec:screenshots}

\begin{figure*}
  \includegraphics[width=0.7\textwidth]{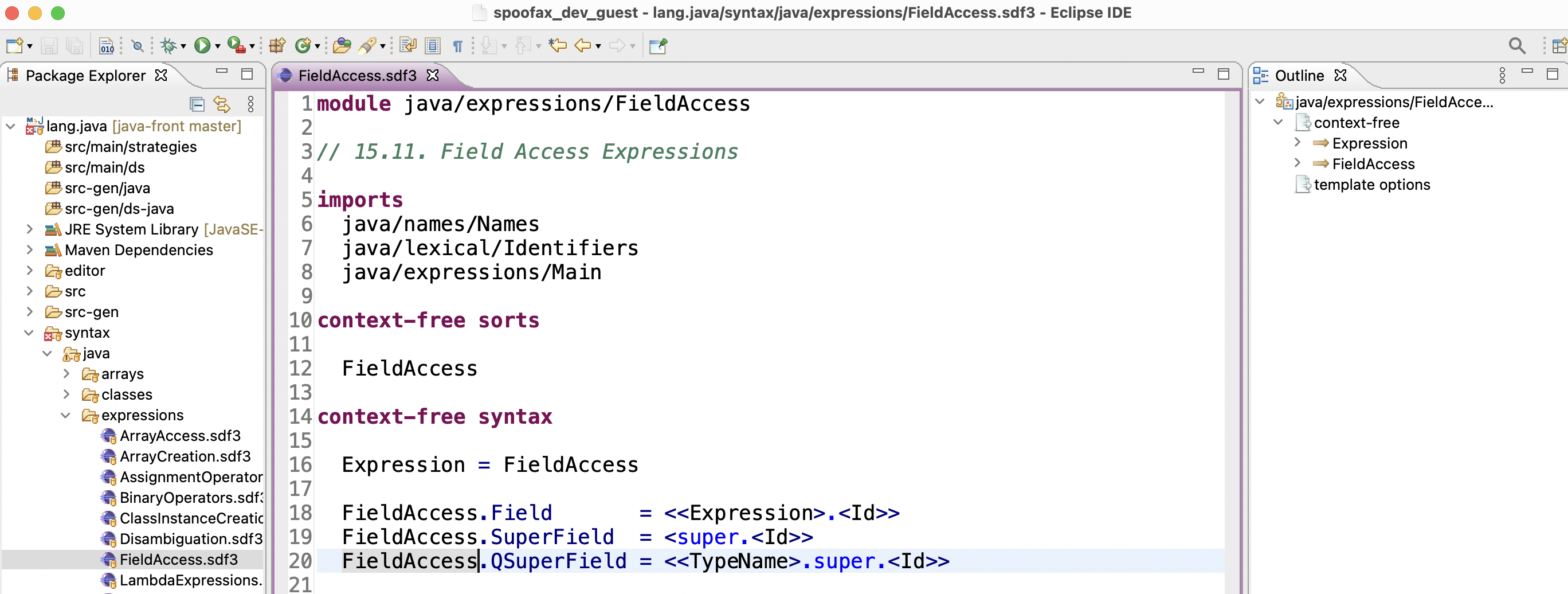}
  \Description{Screenshot of the Spoofax language workbench showing how it looks when editing a file.}
  \caption{Editing a file in the Spoofax language workbench.}
  \label{fig:spoofax}
\end{figure*}

\begin{figure*}
  \includegraphics[width=0.7\textwidth]{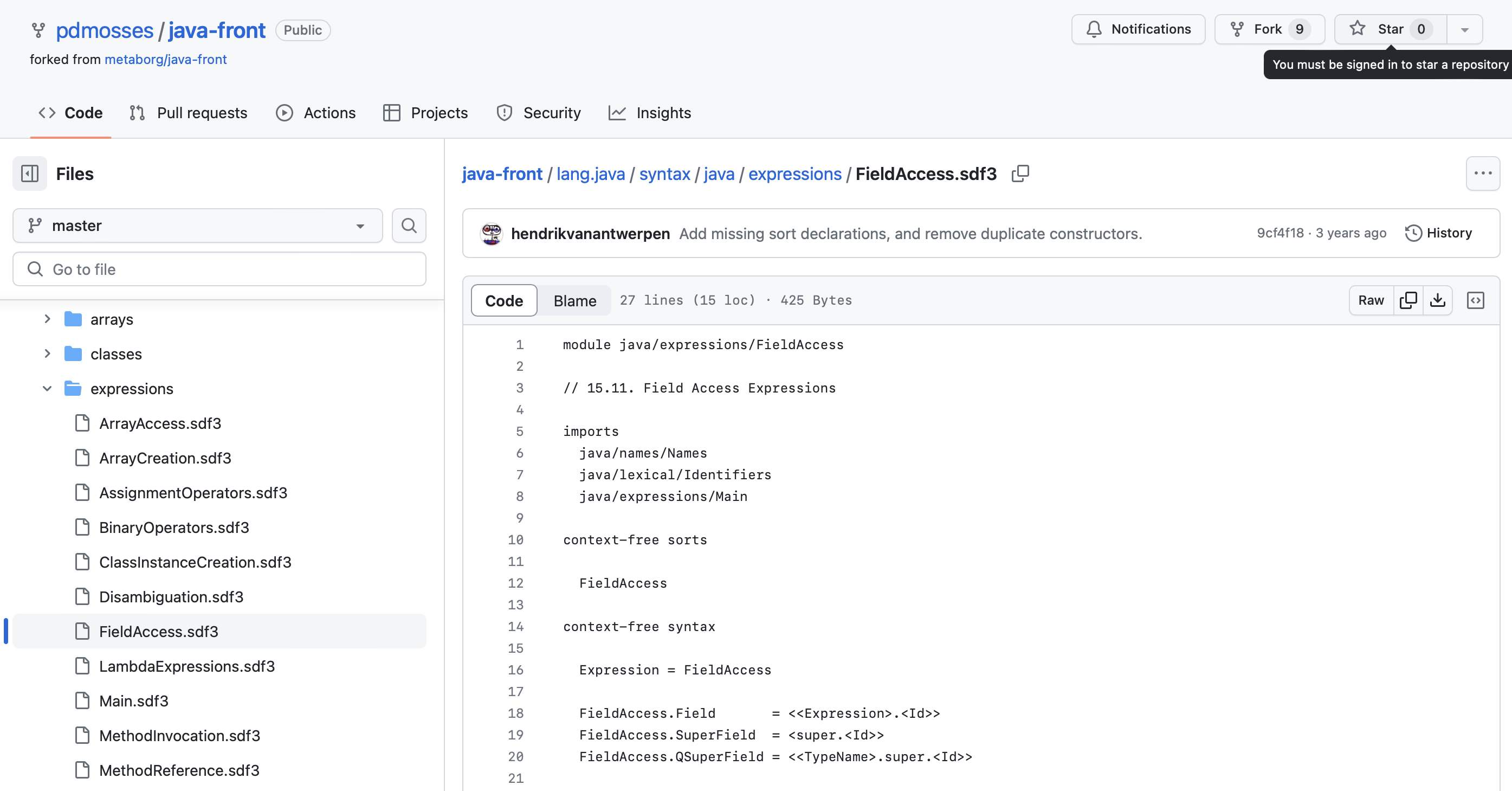}
  \Description{Screenshot of a GitHub repository showing how it looks when browsing the same file as in Figure~\ref{fig:spoofax}.}
  \caption{Browsing the same file in a GitHub repository.}
  \label{fig:github}
\end{figure*}

\begin{figure*}
  \includegraphics[width=0.7\textwidth]{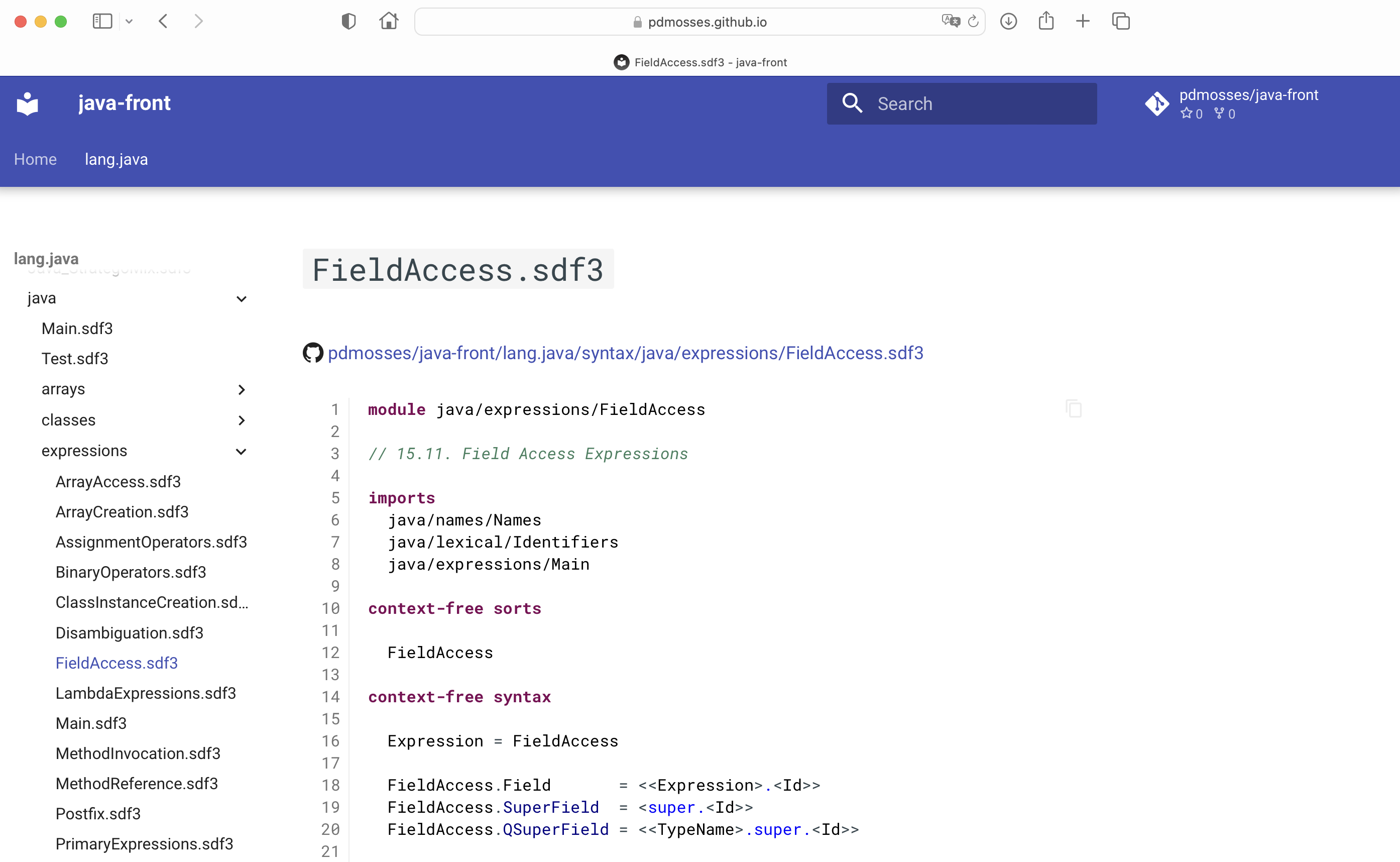}
  \Description{Screenshot of the hyperlinked twin showing how it looks when browsing the same file as in Figure~\ref{fig:spoofax}..}
  \caption{Browsing the same file in the hyperlinked twin.}
  \label{fig:website}
\end{figure*}

The screenshot in Figure~\ref{fig:spoofax} shows how a file from an SDF3 specification of Java looks when editing it in Spoofax. Figure~\ref{fig:github} shows how the same file looks when browsing it on GitHub, and Figure~\ref{fig:website} shows browsing it on the generated hyperlinked twin. Both Spoofax and the hyperlinked twin support name based navigation in SDF3, in contrast to GitHub.

\end{document}